\documentclass[prd,amsmath,amssymb]{revtex4}

\usepackage{graphicx}
\usepackage{bm}

\normalbaselines \topmargin -0.25 truein \oddsidemargin 0.30
truein \evensidemargin 0.30 truein 

\begin{document}

\title{Anisotropic Bianchi-I cosmological model with {\rm SU}(2) - symmetric Yang-Mills field of the magneto-electric type}

\author{Alexander B. Balakin}
\email{Alexander.Balakin@kpfu.ru} \affiliation{Department of General
Relativity and Gravitation, Institute of Physics, Kazan Federal University, Kremlevskaya
str. 18, Kazan 420008, Russia}
\author{Gleb B. Kiselev}
\email{kiselev.gleb.97@gmail.com} \affiliation{Department of General
Relativity and Gravitation, Institute of Physics, Kazan Federal University, Kremlevskaya
str. 18, Kazan 420008, Russia}

\date{\today}

\begin{abstract}
An anisotropic cosmological model with a gauge field, presented by spatially homogeneous {\rm SU}(2)- symmetric Yang-Mills potentials, is considered.
Bianchi-I model with axial symmetry is analyzed; a new exact solution, which describes the early Universe on the stage of domination of the gauge field of the magneto-electric type, is obtained.
\end{abstract}

\maketitle

\section{Introduction}\label{Intro}

For theoreticians working in cosmology the anisotropic homogeneous Bianchi-I model gives a unique space-time platform for testing versions of early Universe expansion. One of the trends in this direction is the analysis of the evolution of the magnetized Universe, i.e., the analysis of models with uniform magnetic field (see, e.g., \cite{BI} for details and references). It is well-known that the symmetry of the Bianchi-I model admits, in particular, the following electromagnetic field configuration: the magnetic field is directed, say, along $0x^3$ axis ($F_{12} = B_0 \neq 0$), and the electric field is parallel to the magnetic one ($F_{03} \neq 0$) (see, e.g., \cite{B2}). The configuration with magnetic and electric fields orthogonal to one another (say, $F_{12}\neq 0$ and $F_{01}\neq 0$) is not compatible with the symmetry of the Bianchi-I model, since the corresponding electromagnetic stress-energy tensor has to have non-admissible non-diagonal component $T_{02}\neq 0$. In addition, even if one deals with the uniform magnetic field only, the potentials for such configuration depend on the spatial coordinates (say, $A_2 {=} \frac12 B_0 x^1$, $A_1 {=} - \frac12 B_0 x^2$). This means that for models, for which the electromagnetic field potentials appear in the stress-energy tensor (e.g., for the Maxwell-Chern-Simons electrodynamics), the model becomes non-homogeneous and thus can not be attributed to the Bianchi-I class.

We study anisotropic cosmological models with the spatially homogeneous Yang-Mills field \cite{BK1,BK2,BK3}; and now we intend to investigate a {\rm SU}(2) analog of the magnetized Universe by introducing the gauge field of the magneto-electric type. The theory of such gauge field deals with the {\rm SU}(2) - symmetric multiplet of vector potentials $A_j^{(a)}$ ($(a)=1,2,3$), and two electrodynamic problems mentioned above can be resolved in elegant way in the framework of this theory. Indeed, one can provide the stress-energy tensor of the Yang-Mills field to be free of non-diagonal components (say, $T_{02}=T_{01}=0$), if the magnetic-like and electric-like components are attributed to different group indices (say, there exist non-vanishing $F_{12}^{(3)}$, $F_{01}^{(2)}$ and $F_{02}^{(1)}$ only). In order to provide the homogeneity of the Yang-Mills field strength $F^{(a)}_{mn}$, which is quadratic in the potentials $A_j^{(b)}$, it is sufficient now to choose these potentials as functions of cosmological time only. This idea is realized and described below.

The work is organized as follows. First, we formulate the complete Einstein-Yang-Mills model with the gauge configuration of the magneto-electric type on the Bianchi-I space-time platform. Then we reduce the total set of master equations for the axially-symmetric submodel and analyze solutions to these reduced equations.

\section{The formalism}

\subsection{Lagrangian of the Einstein-Yang-Mills model and master equations for the gauge and gravity fields }

The action functional of this theory has the standard form
\begin{equation}
S=\int \sqrt{-g} \left\{-\frac{1}{2\kappa}(R+\Lambda) - \frac{1}{4}F^{(a)}_{mn} F_{(a)}^{mn} \right\}  \,.
\label{0}
\end{equation}
Here $R$ is the Ricci scalar, $\Lambda$ is the cosmological constant, $F^{(a)}_{mn}$ is the Yang-Mills field strength given by
\begin{equation}
F^{(a)}_{mn} = \nabla_m A_n^{(a)} - \nabla_n A_m^{(a)} + g f^{(a)}_{\ \ (b)(c)} A_m^{(b)} A_n^{(c)}   \,.
\label{1}
\end{equation}
$A_m^{(a)}$ is the multiplet of the gauge field potentials, $\nabla_m$ is the covariant derivative, $f^{(a)}_{\ \ (b)(c)}$ is the set of group constants, and $g$ is a coupling constant.

Master equations for the Yang-Mills field is known to have the form
\begin{equation}
\nabla_n F^{(a) mn} + g f^{(a)}_{\ \ (b)(c)} A_n^{(b)} F^{(c) mn}  =0   \,,
\label{3}
\end{equation}
and the gravity field equations are
\begin{equation}
R^p_q - \frac12 \delta^p_q (R+2\Lambda) = \kappa T^p_q \,, \quad T^p_q = \frac14 \delta^p_q F^{(a)}_{mn}F_{(a)}^{mn} - F^{(a)pn} F_{(a) qn} \,.
\label{4}
\end{equation}

\subsection{Reduced equations for the Bianchi-I model with {\rm SU}(2) symmetric gauge field}

The metric attributed to the Bianchi-I model is
\begin{equation}
ds^2 = dt^2 - a^2(t){dx^1}^2 - b^2(t)d{x^2}^2 - c^2(t)d{x^3}^2  \,.
\label{Bianchi}
\end{equation}
We assume the speed of light in vacuum is equal to one in the chosen units.
For the {\rm SU}(2) symmetric model $f^{(a)}_{\ \ (b)(c)} {=}\varepsilon^{(a)}_{\ \ (b)(c)}$, where $\varepsilon^{(a)}_{\ \ (b)(c)}$ is the Levi-Civita symbol.
When we are modeling the Yang-Mills field of the magneto-electric type, we assume that there are only two non-vanishing components of the potential:
$A_1^{(2)}(t)$ and $A_2^{(1)}(t)$. Respectively, there are three non-vanishing components of the Yang-Mills field strength:
\begin{equation}
F^{(3)}_{12} = - g A_1^{(2)} A_2^{(1)} \,, \quad  F^{(2)}_{01} = {\dot{A}}^{(2)}_1 \,, \quad  F^{(1)}_{02} = {\dot{A}}^{(1)}_2 \,.
\label{10}
\end{equation}
The dot denotes the derivative with respect to time.
For such gauge field configuration we obtain four non-vanishing components of the stress-energy tensor, and the Einstein equations take the form
\begin{equation}
\frac{\dot{a}}{a} \frac{\dot{b}}{b} + \frac{\dot{a}}{a} \frac{\dot{c}}{c} + \frac{\dot{b}}{b} \frac{\dot{c}}{c} - \Lambda =
\frac{\kappa}{2} \left\{ \frac{1}{b^2}
\left[{\dot{A}}^{(1)}_2\right]^2 + \frac{1}{a^2} \left[{\dot{A}}^{(2)}_1\right]^2  + \frac{1}{a^2 b^2} \left[g A_1^{(2)} A_2^{(1)} \right]^2
\right\} \,,
\label{40}
\end{equation}
\begin{equation}
\frac{\ddot{b}}{b} +  \frac{\ddot{c}}{c} + \frac{\dot{b}}{b} \frac{\dot{c}}{c} - \Lambda =
\frac{\kappa}{2} \left\{ - \frac{1}{b^2}
\left[{\dot{A}}^{(1)}_2\right]^2 + \frac{1}{a^2} \left[{\dot{A}}^{(2)}_1\right]^2  - \frac{1}{a^2 b^2} \left[g A_1^{(2)} A_2^{(1)} \right]^2
\right\}   \,,
\label{41}
\end{equation}
\begin{equation}
\frac{\ddot{a}}{a} +  \frac{\ddot{c}}{c} + \frac{\dot{a}}{a} \frac{\dot{c}}{c} - \Lambda  =
 \frac{\kappa}{2} \left\{  \frac{1}{b^2}
\left[{\dot{A}}^{(1)}_2\right]^2 - \frac{1}{a^2} \left[{\dot{A}}^{(2)}_1\right]^2  - \frac{1}{a^2 b^2} \left[g A_1^{(2)} A_2^{(1)} \right]^2
\right\} \,,
\label{42}
\end{equation}
\begin{equation}
\frac{\ddot{b}}{b} +  \frac{\ddot{a}}{a} + \frac{\dot{b}}{b} \frac{\dot{a}}{a}- \Lambda =
\frac{\kappa}{2} \left\{ - \frac{1}{b^2}
\left[{\dot{A}}^{(1)}_2\right]^2 - \frac{1}{a^2} \left[{\dot{A}}^{(2)}_1\right]^2  + \frac{1}{a^2 b^2} \left[g A_1^{(2)} A_2^{(1)} \right]^2
\right\}
 \,.
\label{43}
\end{equation}
The Yang-Mills potentials $A_1^{(2)}(t)$ and $A_2^{(1)}(t)$ satisfy the master equations
\begin{equation}
\left(\frac{ab}{c}\right)\frac{d}{dt}\left[\left(\frac{bc}{a} \right) \frac{d }{dt} A_1^{(2)} \right] {+} g^2 A_1^{(2)} \left(A_2^{(1)} \right)^2  = 0  \,,
\label{11}
\end{equation}
\begin{equation}
\left(\frac{ab}{c}\right)\frac{d}{dt}\left[\left(\frac{ac}{b} \right) \frac{d }{dt} A_2^{(1)} \right] {+} g^2 A_2^{(1)} \left(A_1^{(2)} \right)^2  = 0  \,.
\label{12}
\end{equation}
The system of master equations contains six equations for five unknown functions; one equation for gravity field can be omitted due to the Bianchi identity.

\section{Analysis of the model with axial symmetry}

\subsection{Reduced system of master equations}

The Bianchi-I model can be indicated as Locally Invariant one, or axially symmetric, when the directions $0x^1$ and $0x^2$ are equivalent and thus  $a(t){=}b(t)$. It is clear from (\ref{41}) and (\ref{42}) that this symmetry is admissible, when
${\dot{A}}^{(1)}_2 {=}\pm {\dot{A}}^{(2)}_1$, or equivalently, $A^{(1)}_2 {=}\pm A^{(2)}_1 {+} {\rm const}$. We assume that the constant is equal to zero and $A^{(2)}_1 {=} A$, thus, we can now work with the following reduced system of three equations for three unknown functions $a(t)$, $c(t)$ and $A(t)$:
\begin{equation}
\left(\frac{\dot{a}}{a}\right)^2 {+} 2 \frac{\dot{a}}{a} \frac{\dot{c}}{c} =  \Lambda {+} \kappa \left(\frac{1}{a^2} {\dot{A}}^2  {+} \frac{g^2}{2a^4}  A^4 \right) \,, \quad
2\frac{\ddot{a}}{a} {+} \left( \frac{\dot{a}}{a} \right)^2 = \Lambda + \kappa \left({-} \frac{1}{a^2}
{\dot{A}}^2 {+} \frac{g^2}{2a^4} A^4 \right)
 \,,
\label{431}
\end{equation}
\begin{equation}
\ddot{A} + \left(\frac{\dot{c}}{c} \right) \dot{A} {+} \frac{g^2}{a^2} A^3  = 0 \,.
\label{111}
\end{equation}
We deal with a set of coupled nonlinear equations, and the study of its solutions requires detailed numerical analysis.
However, in this short communication we consider simple analytical solutions only.

\subsection{Exact solutions}

\subsubsection{Test case, $g=0$}

When $g{=}0$, we deal with the gauge field linear in the Yang-Mills potentials. Equation (\ref{111}) becomes linear, and its first integral is $A(t) {=} \frac{\rm const}{c(t)}$. Clearly, we have to choose vanishing constants, so that ${\dot A}(t) {=}0$, $A(t) {=}0$, providing the magneto-electric configuration to be trivial. The equations (\ref{431}) yield now
\begin{equation}
\frac{\dot a}{a} = \frac{\dot c}{c} = \sqrt{\frac{\Lambda}{3}} \ \Rightarrow \frac{a(t)}{a(t_0)} = \frac{c(t)}{c(t_0)} = e^{\sqrt{\frac{\Lambda}{3}}(t-t_0)}\,.
\label{66}
\end{equation}
In other words, the space-time without gauge field of the magneto-electric type has to be the isotropic Friedmann one, as it should be.

\subsubsection{Example of exponential solution}

We search for the unknown functions in the exponential form
\begin{equation}
a(t) = a(t_0) \ e^{h(t-t_0)} \,, \quad c(t) = c(t_0) \  e^{\mu (t-t_0)} \,, \quad A(t) = \nu a(t)  \,.
\label{88}
\end{equation}
Equations (\ref{431}), (\ref{111}) give three relationships, which link four constant parameters $h, \mu, \nu, \Lambda$:
\begin{equation}
h^2 {+} 2 h \mu =  \Lambda {+} \kappa \left( \nu^2 h^2 {+} \frac{g^2 \nu^4}{2} \right) \,,
\quad
3h^2 = \Lambda {+} \kappa \left(- \nu^2 h^2  {+} \frac{g^2 \nu^4}{2} \right)
 \,, \quad
 \nu h^2 {+} \nu \mu h {+} g^2 \nu^3  = 0 \,.
\label{511}
\end{equation}
These relationships can be rewritten in the form of parametric representation:
\begin{equation}
\mu = h \frac{(g^2-\kappa h^2)}{(g^2+\kappa h^2)} \,,\quad
\nu^2 = - \frac{2h^2}{(g^2+\kappa h^2)}  \,, \quad
\Lambda = h^2 \left[1 + \frac{2g^4}{(g^2+\kappa h^2)^2} \right]
  \,.
\label{77}
\end{equation}
The parameter $h$ describes the rate of a Universe expansion in the plane $x^10x^2$; the parameter $\mu$ relates to the rate of evolution in the direction $0x^3$. The parameter $\nu= \pm i \sqrt{\frac{2h^2}{(g^2+\kappa h^2)}}$ is a factor forming the amplitude of the gauge field potentials.  The Yang-Mills field strength components are of the form
\begin{equation}
F^{(3)}_{12} = - g \nu^2 a^2 = \frac{2g h^2}{g^2 {+}\kappa h^2} a^2(t_0) e^{2h(t{-}t_0)} \,, \quad F^{(2)}_{01} = \pm F^{(1)}_{02} = \nu h a(t)\,.
\label{444}
\end{equation}
Respectively, the physical components of the gauge analogs of the magnetic and electric fields, $B$ and $E$, are the constant quantities:
\begin{equation}
B \equiv \sqrt{F^{(3)}_{12}F_{(3)}^{12}} = \left|g \nu^2 \right| = \frac{2g h^2}{g^2 {+}\kappa h^2}  \,,
\quad E \equiv \sqrt{{-} F^{(2)}_{01} F_{(2)}^{01}}= \sqrt{{-} F^{(1)}_{02} F_{(1)}^{02}}= |\nu| h \,.
\label{445}
\end{equation}
It is interesting to mention, that in case of Bianchi-I model with U(1) magnetic field the component $F_{12}$ is constant \cite{B2}, but the physical component B behaves as $B(t)=B_0/a^2$. However, in {\rm SU}(2) model the situation is opposite: the component $F^{(3)}_{12}$ depends on time (see (\ref{444})), but the physical component happens to be constant (\ref{445}).

\section{Outlook}

In this short communication based on exact solutions (\ref{88}), (\ref{77}) of the master equations (\ref{431}), (\ref{111})  we discussed the structure of the {\rm SU}(2) symmetric Yang-Mills field configuration of the magneto-electric type in the context of anisotropically expanding Bianchi-I space-time platform with axial symmetry. In the next paper we hope to supplement this analytical investigation by detailed numerical analysis of the model.



\end{document}